# Local Nanoscale Defective Phase Impurities Are the Sites of Degradation in Halide Perovskite Devices

Stuart Macpherson[1†], Tiarnan A. S. Doherty[1†], Andrew J. Winchester[2], Sofiia Kosar[2], Duncan N. Johnstone[3], Yu-Hsien Chiang[1], Krzystof Galkowski[1], Miguel Anaya[1], Kyle Frohna[1], Affan N. Iqbal[1,4], Bart Roose[4], Zahra Andaji-Garmaroudi[1], Paul A. Midgley[3], Keshav M. Dani*[2], Samuel D. Stranks*[1,4]

1. Department of Physics, Cavendish Laboratory, University of Cambridge, Cambridge, UK
2. Femtosecond Spectroscopy Unit, Okinawa Institute of Science and Technology, Onna-son, Okinawa, Japan
3. Department of Materials Science and Metallurgy, University of Cambridge, Cambridge, UK
4. Department of Chemical Engineering and Biotechnology, University of Cambridge, Cambridge, UK

† These authors contributed equally.

*sds65@cam.ac.uk, *kmdani@oist.jp

**Halide perovskites excel in the pursuit of highly efficient thin film photovoltaics and light emitters[1]. Power conversion efficiencies have reached 25.5% in single junction and 29.5% in tandem halide perovskite/silicon solar cell configurations by employing mixed composition perovskite absorber layers, with tandems now exceeding record single junction silicon cells[2,3]. Operational stability of perovskite solar cells remains a barrier to their commercialisation[4], yet a fundamental understanding of degradation processes, including the specific sites at which failure mechanisms occur, is lacking. Recently, we reported that performance-limiting deep sub-bandgap states appear in nanoscale clusters at particular grain boundaries in state-of-the-art $Cs_{0.05}FA_{0.78}MA_{0.17}Pb(I_{0.83}Br_{0.17})_3$ (MA=methylammonium, FA=formamidinium) perovskite films[5]. Here, we combine multimodal microscopy to show that these very nanoscale defect clusters, which go**

**otherwise undetected with bulk measurements, are sites at which degradation seeds. We use photoemission electron microscopy to visualise trap clusters and observe that these specific sites grow in defect density over time under illumination, leading to local reductions in performance parameters such as luminescence[6]. Scanning electron diffraction measurements reveal concomitant structural changes at phase impurities associated with trap clusters, with rapid conversion to metallic lead through iodine depletion, eventually resulting in pinhole formation. By contrast, illumination in the presence of oxygen – a mechanism known to passivate trap states[7,8] – reduces defect densities and reverses performance degradation at these local clusters, where phase impurities instead convert to amorphous and electronically benign lead oxide. Remarkably, defect densities at these sites can be modulated orders of magnitude by cycling between oxygen and oxygen-free environments under illumination, highlighting their dynamic nature. Our work shows that the trapping of charge carriers at sites associated with phase impurities, itself reducing performance, catalyses redox reactions that compromise device longevity. Importantly, we reveal that both performance losses and intrinsic degradation can be mitigated by eliminating these defective clusters.**

Thin films of mixed composition $Cs_{0.05}FA_{0.78}MA_{0.17}Pb(I_{0.83}Br_{0.17})_3$ perovskite, representative of state-of-the-art single junction[9] and top-cell tandem solar cell absorbers[10], were solution-processed on indium-tin-oxide (ITO)-coated glass substrates and SiN transmission electron microscopy (TEM) grids. Single-junction solar cell devices constructed from these same absorber layers sandwiched between electron and hole collecting contacts exhibit excellent photovoltaic performance (see Methods for fabrication details and Extended Data Figure 1 for material and device characterisation). We employ photoemission electron microscopy (PEEM) to spatially and energetically characterise occupied valence band and sub-bandgap states on the surface of the absorber films (see Methods)[5]. Figure 1a shows the photoemission intensity from a distribution of surface trap clusters on the as-prepared sample. The clusters range in size from tens to hundreds of nanometres (mean diameter = 140 nm, standard deviation = 70 nm), occupying ~14% of the film surface (Extended Data Figure 2). Furthermore, the clusters vary in defect density (following the PEEM intensity), shape and distribution across the film. The spatially averaged photoemission spectrum (PES; see Methods) of the as-prepared sample shown in Figure 1b (0 hours) reveals a broad distribution of occupied states extending into the sub-gap region, up to the Fermi level (Extended Data Figure 3a and b). We have previously identified these clusters as traps for hole charge carriers and that they are responsible for non-radiative recombination losses in these films[5].

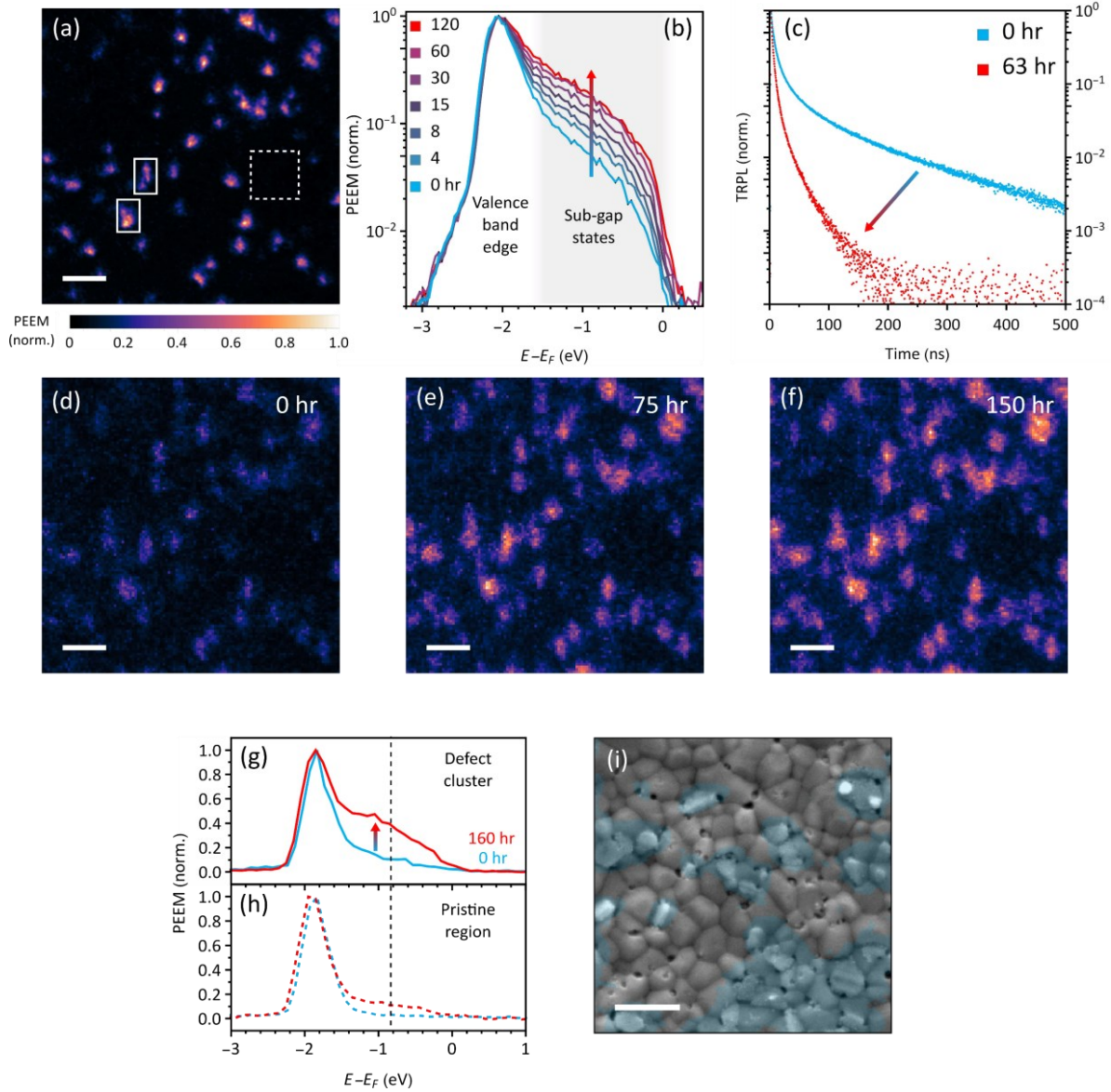

**Figure 1: Trap clusters are sites at which local degradation seeds. (a)** Normalised PEEM image of sub-bandgap photoemission from trap clusters on the surface of a $Cs_{0.05}FA_{0.78}MA_{0.17}Pb(I_{0.83}Br_{0.17})_3$ perovskite film (4.65 eV probe). **(b)** Spatially averaged photoemission spectra at stages of *in situ* solar-equivalent illumination dose (λ = 400 nm; see Methods and Extended Data Table 1). **(c)** Time-resolved photoluminescence decay before and after 63 hours of solar-equivalent illumination. Spatially resolved PEEM intensity at sub-bandgap energy $E-E_F$ = -0.83 ± 0.15 eV (vertical line in panel g and h), after **(d)** 0 hours, **(e)** 75 hours and **(f)** 150 hours of *in situ* solar-equivalent illumination. Local photoemission spectra from **(g)** isolated trap clusters (solid box in panel a) and **(h)** a pristine sample region (dashed box), before (blue) and after (red) 160 hours of solar-equivalent illumination. **(i)** Thin film morphology measured in a scanning electron microscope after 150 hours of solar-equivalent illumination. Blue overlay: PEEM mask showing highest intensity of sub-bandgap photoemission

(50% threshold) from the pristine sample before illumination. Light exposure was carried out under high vacuum conditions ($10^{-10}$ Torr in PEEM and $10^{-5}$ Torr in PL). All scale bars are 500 nm.

To understand how these nanoscale trap clusters evolve over time under operation, we monitor the photo-induced evolution of the electronic structure through PEEM measurements under continuous illumination. Here, we illuminate the film *in situ* with continuous laser illumination at 400 nm and benchmark the excitation density and duration to equivalent solar illumination dose (see Methods and Extended Data Table 1). We note that the low-dose PEEM probe does not induce sample changes (see Methods and Extended Data Figure 3)[5]. As shown in the spatially averaged PES in Figure 1b, we observe an increase in photoemission intensity at energies in the sub-bandgap regime, down to 1.5 eV below the Fermi level, with increasing time under illumination. The density of sub-bandgap trap states approximately doubles within the first 20 hours of solar-equivalent dose, followed by a subsequent slower growth regime (Figure 1b and Extended Data Figure 3). The absolute increase in sub-bandgap states is greatest in the shallow region close to the valence band edge but interestingly the greatest relative growth comes from deeper within the sub-bandgap regime (see Extended Data Figure 3). The growth of this sub-bandgap distribution also correlates with a large decrease in the photoluminescence (PL) lifetime (Figure 1c) and drop in PL intensity of ~80% after illumination for 63 hours under the same conditions (Extended Data Figure 4). This is consistent with an increased density of non-radiative recombination centres observed after light exposure[11]; a similar drop in performance is reported in full devices after long periods of illumination[12].

The progression of Figures 1e, f and g show the PEEM maps corresponding to the sub-gap trap regime (energetically filtered at $E - E_f = -0.83 \pm 0.15$ eV, vertical line in figure 1g and h) at respective time snapshots of 0, 75 and 150 hours of solar-equivalent illumination dose. Importantly, the spatial distribution of trap clusters after illumination is largely unchanged;

there is negligible generation of new clusters within pristine sample regions that do not initially have detectable trap densities (see Extended Data Figure 5 for logarithmically scaled images). Strikingly, there is significant growth of the PEEM intensity within the sub-bandgap regime for the initially defective sites, while the more pristine sample regions experience minimal change, insinuating that there are far fewer trap states forming in the initially pristine regions. This observation is supported by the increased heterogeneity in the photoemission distribution after 150 hours solar-equivalent dose (see Extended Data Figure 5). Furthermore, these results are consistent with confocal PL maps taken *in situ* under similar illumination conditions, which reveal that the relative standard deviation of the spatial luminescence doubles, indicating an increase in PL heterogeneity, with regions that are initially darker in luminescence showing the most significant drop in PL signal (Extended Data Figure 4). The changes in sub-gap states at local sites are further probed by analysing the PEEM energy spectra of discrete locations on the sample (see Methods). The spectral changes before (0 h) and after (160 h) illumination at trap clusters (Figure 1g; highlighted by the solid boxes in Figure 1a) compared to a pristine sample region with no detectable initial sub-bandgap PEEM intensity (Figure 1h; dashed box in Figure 1a) show that the sub-bandgap photoemission signal increases almost exclusively at existing trap clusters during the light exposure. We thus conclude that the photo-induced generation of trap clusters is seeded at existing sites and occurs selectively at defect-rich regions on the sample surface. In absolute terms, the emergence of sub-bandgap trap states is accelerated at initially defective locations.

Scanning electron microscopy (SEM) images of the films (Figure 1i) after illumination reveal extensive pinhole formation at what appear to be grain boundaries. Such morphological imperfections are not present before light exposure (Extended Data Figure 1). Strikingly, the overlay of the initial PEEM distribution of the film (blue in Figure 1i) reveals that these morphological changes occur primarily at the same regions which display the greatest initial trap density, and which is associated with the lowest optoelectronic performance. A similarly clear relationship is observed between regions that exhibit large growth in PEEM signal and local morphological change (Extended Data Figure 3), showing that pinholes align with sites of increased trap density (PEEM intensity), even if the trap clusters are not detectable initially. Such pinhole distributions are also observed in full solar cell devices from similar compositions that have degraded after 500 hours of operation under 1 sun illumination.[13] This confirms that our measurements are probing the key underlying degradation pathways of devices which are limited by intrinsic properties of the absorber layer, albeit the effects are slower in full devices due to the protecting impact of contact layers[14] and the lower carrier density induced at maximum power point compared with open circuit voltage. Connecting the presence and formation of traps, with local material decomposition, establishes a direct link between charge trapping and long-term deterioration of device performance.

To gain structural insight into the early-stage mechanisms driving the observed optoelectronic changes and material decomposition, we performed scanning electron diffraction (SED) measurements on a $Cs_{0.05}FA_{0.78}MA_{0.17}Pb(I_{0.83}Br_{0.17})_3$ perovskite sample before and after illumination for 1-hour solar illumination-equivalent (see Extended Data Figure 6 for corresponding PL measurements during the light exposure). We preclude the influence of electron beam-induced degradation on the interpretation of our results in these low dose measurements (see Methods and Extended Data Figure 6). Diffraction sum images revealing the film morphology of an area of the sample before (Figure 2a) and after illumination (Figure 2b) reveal observable changes only in local areas of the sample, consistent with the results from Figure 1.

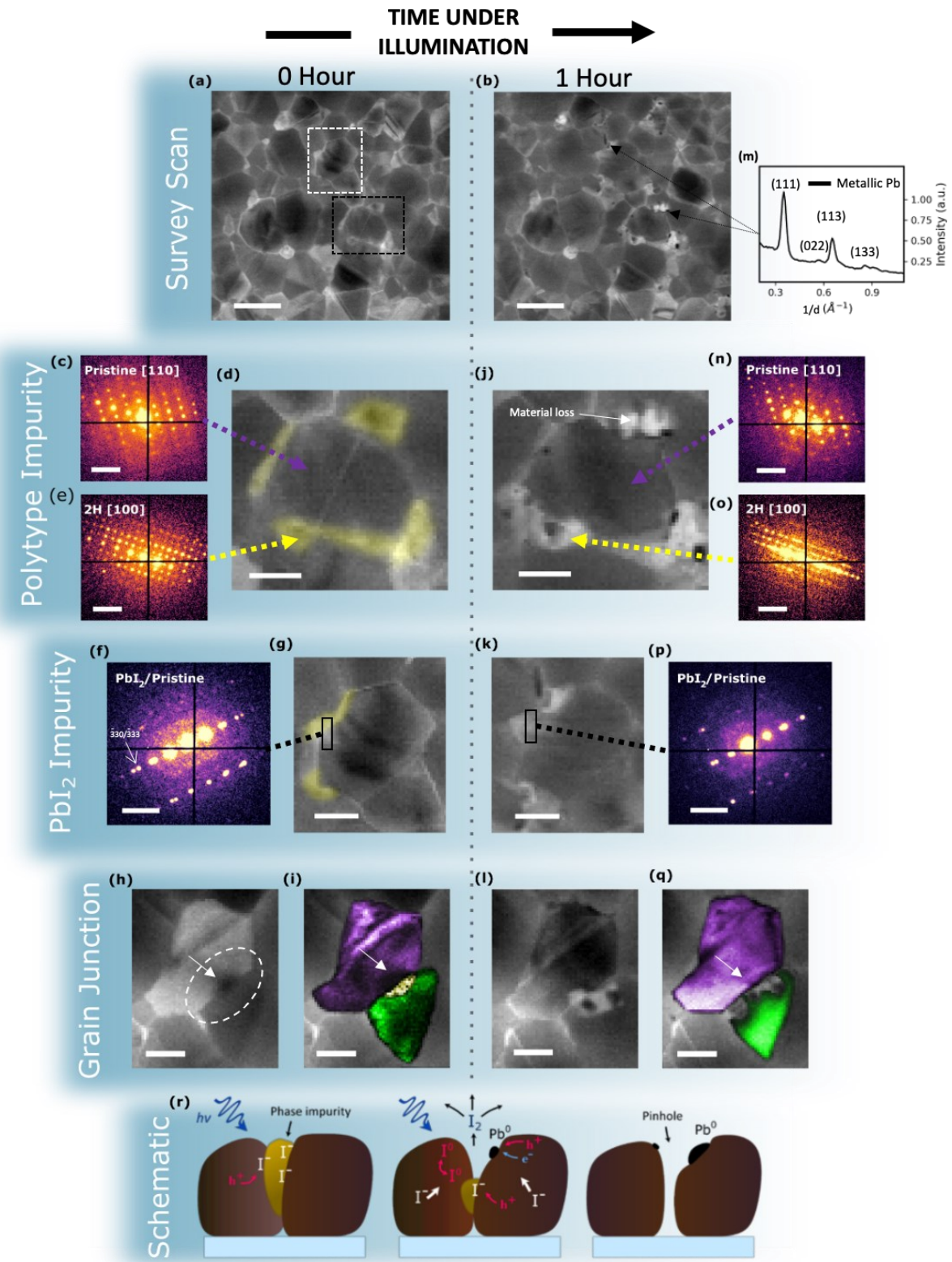

**Figure 2: Light-induced degradation occurs at phase impurities in $Cs_{0.05}FA_{0.78}MA_{0.17}Pb(I_{0.83}Br_{0.17})_3$ thin films, ascertained by monitoring structural changes after illumination.** Diffraction sum images from scanning electron diffraction (SED) measurements showing a survey scan of a region of a $Cs_{0.05}FA_{0.78}MA_{0.17}Pb(I_{0.83}Br_{0.17})_3$

thin film **(a)** prior to illumination and **(b)** on the same area following 1 hour of solar illumination-equivalent *in vacuo* (< $10^{-6}$ mbar). Distinct changes after illumination are visible in some regions of the film. Scalebar is 300 nm. **(c)** Diffraction pattern indexed to the [110] zone axis of a pseudo-cubic perovskite ('pristine') extracted from the grain indicated by a purple arrow in (d). **(d)** Diffraction sum image extracted from SED data showing a region (black box region of interest in (a)) prior to illumination. 2H hexagonal regions surrounding the grain are coloured in yellow. **(e)** Diffraction pattern extracted from the yellow region indicated in (d) indexed to the [100] zone axis of a 2H hexagonal perovskite. **(f)** Diffraction pattern extracted from the region indicated in (g) which encompasses a pristine perovskite grain and epitaxially aligned grain boundary $PbI_2$. One set of doubled diffraction spots is indexed to pristine perovskite (-333) and $PbI_2$ (-330). **(g)** Diffraction sum image extracted from SED data showing a region (white box region of interest in (a)) revealing a pristine perovskite grain surrounded by epitaxially aligned $PbI_2$ phase impurities that are coloured in yellow. **(h)** Diffraction sum image of a grain junction. The white arrow indicates a spatial variation in diffraction contrast within the grain. **(i)** Virtual dark field (VDF) images overlayed on a diffraction sum image from the same region as (h) revealing a small phase impurity (yellow) at the interface of two grains (purple and green). **(j)** Diffraction sum image of the same region of the film as (d) following 1 hour of solar illumination-equivalent exposure. **(k)** Diffraction sum image of the region of the film indicated in (g) following 1 hour illumination. **(l)** Diffraction sum image of the region of the film indicated in (h) following 1 hour of illumination. **(m)** Average azimuthally integrated diffraction pattern extracted from a number of metallic Pb precipitates. **(n)** Diffraction pattern extracted from a pristine perovskite grain (indicated by the purple arrow in (j)) following illumination. **(o)** Diffraction pattern extracted from 2H hexagonal phase impurity following 1 hour of solar-equivalent illumination. **(p)** Diffraction pattern extracted from the black box region in (k) showing the presence of epitaxially aligned pristine perovskite and $PbI_2$ following illumination. **(q)** VDF images overlayed on a diffraction sum image from the same region as (l) following 1 hour of solar illumination-equivalent exposure. **(r)** Schematic illustrating progression of light-induced material degradation at a local phase impurity. Hole photocarriers oxidise iodide defects triggering the reaction of neutral iodine species to form molecular iodine which is lost as a gas. Residual metallic $Pb^0$ clusters form adjacent to the degraded site and further quench photocarriers. Iodide ions move to towards the film surface, fuelling further degradation which results in eventual pinhole formation. Scalebars for (c), (e), (g), and (n) – (p) is 0.5-$A^{-1}$; for (d), (g) – (k), (l) and (q) is 100 nm.

To elucidate these changes in local structural properties, we first closely examine a number of these sites of degradation prior to illumination to establish the initial local structural landscape.

In Figure 2c, we show the diffraction pattern extracted from the grain indicated in Figure 2d (purple arrow) before illumination. This diffraction pattern can be indexed to a pseudo-cubic perovskite structure with a lattice parameter of 6.3 Å that is oriented near the [110] zone axis, which we herein refer to as 'pristine'.[5] Surrounding this pristine grain are phase impurities (coloured yellow in Figure 2d) that are assigned to inclusions, tens of nanometres in size, of a 2H hexagonal perovskite polytype oriented near the [100] zone axis (see Extended Data Figure 6 and Methods), as revealed by the diffraction pattern extracted from this phase impurity (Figure 2e). We also observe other phase impurity species associated with subsequent changes under illumination: In Figure 2f, we show the combined diffraction pattern of pristine perovskite and epitaxially aligned $PbI_2$ impurities acquired from the region indicated in Figure 2g, consistent with recent reports of $PbI_2$/perovskite intergrowth in pseudo-cubic perovskites;[15] the epitaxially aligned $PbI_2$ is denoted in yellow in Figure 2g (see Extended Data Video 1 and Extended Data Figure 6 for further analysis). Finally, in Figure 2h, we highlight an interface in the diffraction sum image that appears to be between two distinct grains (white ellipse, Figure 2h). Reconstruction of virtual dark field (VDF) images that plot the spatially resolved intensity of selected Bragg peaks from the SED data (Extended Data Figure 7 for diffraction patterns, crystal assignments and analysis) reveals the presence of an additional small crystalline grain (denoted yellow in Figure 2i) that is nestled between two perovskite grains (purple and green in Figure 2i). This small grain can be assigned to either a hexagonal perovskite polytype or $PbI_2$; although unambiguous assignment is not possible, it is in any case again a phase impurity (see Extended Data Figure 7 and Methods for discussion). Notably, variations in diffraction contrast indicated by white arrows in Figures 2h and 2i are observed, which indicates that the purple pristine grain (Figure 2i, and Extended Data Figure 7) bends out of the Bragg condition as it approaches the phase impurity, potentially indicating a strained contact point at the interface between the pristine grain and the phase impurity (see Extended Data Figure 7 and Methods for detailed analysis).

Following illumination, we observe substantial changes at these sites of phase impurities concomitant with decreases in PL (Extended Data Figure 7). Dark spots appear in the diffraction sum images at each of these sites (Figures 2b, j, k, l) which we identify as precipitates of crystalline metallic lead formed during illumination; two examples are indicated by arrows in Figure 2b. We confirm this assignment in Figure 2m by spatially averaging and azimuthally integrating the 2D diffraction patterns extracted from a number of these dark metallic lead spots across multiple samples exposed to the same illumination conditions (see Methods and Extended Data Figure 8). Surrounding, or directly adjacent to, the dark metallic lead precipitates at each of the sites (Figures 2b, j, k, l) are regions of light diffraction contrast that we ascribe to material loss, and which we discuss in further detail below; an example is indicated by the white arrow in Figure 2j. The crystal structure of the pristine perovskites following illumination remains largely unchanged (Figure 2n) barring a slight reorientation of the grain, which we observe ubiquitously across the film including at regions away from degradation sites (see Extended Data Figure 9 for patterns at regions away from degradation sites). By contrast, the crystal structure of the phase impurities is substantially altered: In a diffraction pattern extracted from the 2H hexagonal phase impurity following illumination (Figure 2o), we observe a large amount of streaking in the [002] direction, which we attribute to either coexistence of multiple degradation products together with the original 2H lattice, and/or a high density of new structural defects, such as vacancies, consistent with our assignment of material loss in regions of light diffraction contrast. In some regions of the 2H hexagonal impurity phases, we also observe a crystal phase that most closely resembles a $PbI_2$ 4H polytype[16] as another degradation product, in addition to the metallic $Pb^0$ (Extended Data Figure 9). At the $PbI_2$ phase impurity site, a small degree of epitaxially aligned $PbI_2$ is still present after illumination (Figure 2p, see Extended Data Figure 9 for examples of entire grains of PbI$_2$ also changing under illumination). However, at the 'grain-junction' impurity site (Figure 2h, i, l, q) the small phase impurity that was present at the boundary of the pristine perovskite prior to illumination (yellow grain image in Figure 2i) has disappeared entirely, suggesting that it has degraded into an amorphous material that no longer produces a Bragg diffraction spot. Accompanying the loss of this small grain, there is an homogenisation of the diffraction contrast within the purple pristine grain (Figure 2q). We interpret this as a relaxation of the underlying crystal structure of the pristine grain due to changes in its local structural environment, i.e., the removal of the phase impurity. Surprisingly, examining the VDF extracted from the grain on the other side of the junction (green grain, Figure 2q) reveals that even the pristine material has started to 'etch' away from where the yellow phase impurity was originally located. This points to a loss in crystallinity in the green grain (Figure 2q) and leads to an important conclusion that these phase impurities can seed further degradation in locally adjacent regions. Thus, extensive material loss at the sites of these impurities seeds the same in directly adjacent regions and leads to the eventual formation of pinholes in the perovskite absorber film (see Extended Data Figure 8), in line with the morphological deterioration at the initially trap-rich clusters highlighted in Figure 1i. By contrast, the pristine sample regions away from these impurity sites remain mostly unaffected. Combined, these measurements directly show the crucial role of phase impurities in seeding the light-induced degradation of halide perovskites.

We propose a global mechanism by which light-induced degradation seeds at local phase impurities, in Figure 2r. These impurities appear in our PEEM measurements as clusters of sub-bandgap states which become increasingly dense over time under illumination and are associated with eventual material loss (cf. Figure 1). We propose that polytype impurities contain high densities of defects such as interstitial iodide, which act as traps for photo-excited hole carriers[5,17] and facilitate non-radiative recombination. Trapped holes oxidise iodine interstitial defects, providing fuel for local bimolecular reactions of neutral iodide species (Figure 2r)[18–20]. The resulting molecular iodine is lost as a gas at the material surface, explaining the expedited reduction in crystallinity and loss of material at local impurity sites. Nanoscopic lead formations are a product of the extensive iodine loss at the material surface: as iodine escapes, more iodide moves into the region to compensate through photo-induced and defect-mediated ion migration,[21,22] generating more uncoordinated $Pb^{2+}$ in the vicinity, which is in turn susceptible to reduction to metallic $Pb^0$ by photo-excited electrons. Within locally iodine-rich regions, $I_2^-$ radicals and triiodide ions ($I_3^-$) may additionally catalyse the degradation of adjacent perovskite material by reducing the organic cation[23,24], exacerbating the local material instability which promotes eventual pinhole formation (Figure 2r). Our results are also consistent with the reported photolysis of highly defective[16,25] $PbI_2$, which degrades via trap-assisted redox reactions to form $Pb^0$ and molecular iodine ($I_2$; see schematic in Extended Data Figure 8)[26,27]. $Pb^0$ accumulations act as an additional quenching pathway for photo-excited carriers[28,29], further increasing the local sub-gap density of states (as seen in PEEM). Furthermore, we detect light-induced segregation of iodine and bromine rich material (Extended Data Figure 8) which will lead to local bandgap variations and the preferential accumulation of carriers on lower-bandgap, iodide-rich sites[30] (consistent with red-shifted photoluminescence; Extended Data Figure 6), further concentrating photo-chemically driven redox reactions at these locations. Over time under illumination, the combination of interstitial iodine accumulation, $Pb^0$ formation and increased disorder explains the local growth in the sub-gap trap densities, and subsequent pinhole formation at phase impurities. Crucially, comparatively low defect populations restrict these unwanted photochemical processes in pristine sample regions. Carrier traps themselves are therefore the key drivers for local film degradation.

To further test the hypothesis that carrier trap clusters, relating to local phase impurities, are responsible for seeding degradation, we dynamically modulate the trap density through passivation by varying atmospheric conditions during illumination. Illuminating films over time in vacuum leads to a rapid drop in PL intensity and lifetime (red line in Figure 3a and 3b), but then further illumination of the same sample in an oxygen (ambient air) environment leads to a dramatic increase in PL intensity by multiple orders of magnitude and an improved lifetime beyond that of the unexposed film.[8] Such optoelectronic enhancements in the presence of oxygen and light have been attributed to surface trap passivation through redox reactions between oxygen, ionic species and photo-excited electrons,[11,18,31] though a complete nanoscale understanding of these effects has remained elusive. We compare the trap cluster distributions ascertained from PEEM measurements of an as-prepared sample (Figure 3c) and of the same region following *in-situ* illumination in vacuum (Figure 3d), and then following *ex-situ* illumination in the presence of oxygen (Figure 3e; see Methods and Extended Data Table 1). Light exposure in vacuum results in locally enhanced sub-bandgap photoemission at increasingly defective clusters (cf. Figure 1). Strikingly, after subsequent illumination of the same film in the presence of oxygen the sub-bandgap photoemission intensity is drastically reduced on those same sites. Figure 3f shows how the distribution of local trap densities changes with each treatment, and that upon illumination in oxygen the mean sub-bandgap photoemission intensity decreases by a factor of ~8 from the values after vacuum light exposure, corresponding to a factor of ~6 drop from the initial values in the as-prepared film before any illumination. PL enhancements and trap-density (PEEM) reductions are also observed in as-prepared films that are only illuminated in oxygen (see Extended Data Figure 10). Photoemission spectra show the valence band edge to be unchanged while the sub-bandgap photoemission is significantly quenched, dropping by an order of magnitude in the mid-gap regime (Extended Data Figure 10). A small work function increase can be observed from the narrowing of the photoemission spectrum and the shift of the cut-off to higher energy.[32] Changes in trap-related photoemission are not spatially uniform: not only is the absolute reduction in sub-bandgap photoemission intensity (trap density) greater at the clusters initially of highest intensity, the changes result in a more uniform trap intensity distribution across the sample. We thus conclude that the most dramatic changes when illuminating in oxygen occur at the regions associated with the highest density of traps (PEEM intensity), again highlighting the critical photo-activity at these defective sites.

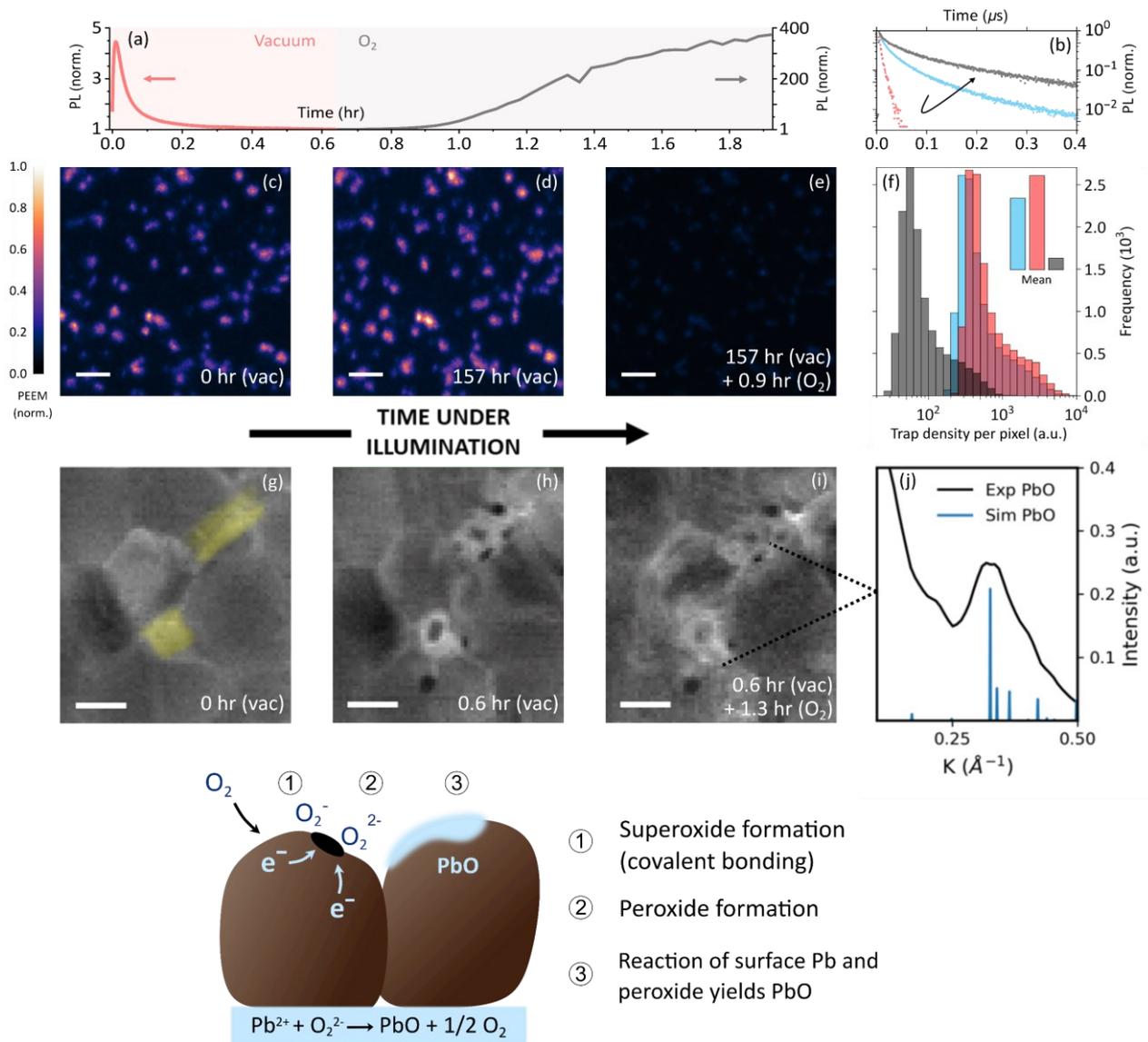

**Figure 3: Trap clusters and local phase impurities are dynamically modified through light and atmosphere, in turn modulating the optoelectronic response. (a)** Spectrally integrated photoluminescence intensity of a $Cs_{0.05}FA_{0.78}MA_{0.17}Pb(I_{0.83}Br_{0.17})_3$ perovskite film over time under solar illumination conditions, *in vacuo* (red), followed by the PL evolution of the same sample under solar illumination conditions in an oxygen environment (black). **(b)** Time-resolved photoluminescence decay curves for a $Cs_{0.05}FA_{0.78}MA_{0.17}Pb(I_{0.83}Br_{0.17})_3$ perovskite film as-prepared (blue), after illumination for 70-hours equivalent solar dose in vacuum (red) and after subsequent illumination for 1.2 hours equivalent solar dose in an oxygen environment (black). Maps of sub-bandgap photoemission of **(c)** an as-prepared film, **(d)** the same region after exposure to illumination in vacuum (157 hours solar equivalent dose) and **(e)** the same region after exposure to illumination in an oxygen-rich environment (0.9 hour solar equivalent dose). Scalebars are 500 nm. **(f)** Histograms showing distribution of trap density (sub-bandgap PEEM) across images in panels c-e. Inset: Relative mean intensity of each distribution. **(g)** Diffraction sum image from with phase impurity regions shaded in yellow. **(h)** Diffraction contrast image after exposure to illumination for 0.6 hours solar-equivalent dose, in vacuum. **(i)** Diffraction contrast image after exposure to

illumination for 1.3 hours solar-equivalent dose in an oxygen-rich environment (see Methods; scalebars = 100 nm). **(j)** Azimuthally integrated diffraction pattern from amorphous regions of panel f, consistent with primary bond lengths of PbO. Simulated XRD pattern of PbO is plotted in blue. **(k)** Schematic illustrating the conversion of surface lead species to PbO upon illumination in an $O_2$-rich environment.

To reveal the structural evolution associated with cycling the optoelectronic properties under illumination in different atmospheres, we perform SED measurements to monitor the same samples shown in the PL time series of Figure 3a. In Figure 3g, we show a summed diffraction image of a defective region of the perovskite film prior to illumination, with 2H hexagonal phase impurities highlighted in yellow (see Extended Data Figure 9 for indexation). Following exposure to illumination in vacuum (0.6 hour solar equivalent illumination), metallic $Pb^0$ is formed at these phase impurities as degradation is triggered at these sites (Figure 3h), concomitant with the dramatic drop in PL. However, under illumination in oxygen (coinciding with the substantial increase in PL), the metallic $Pb^0$ and residual phase impurities at these degraded sites are transformed to an amorphous material (Figure 3g, Extended Data Figure 9). By azimuthally integrating the ring pattern obtained from the amorphous material (Figure 3h), we find that the primary bond lengths of the amorphous material are consistent with PbO, which forms specifically at these dynamic impurity sites. Similar examples of PbO formation are observed at sites initially corresponding to $PbI_2$ (see Extended Data Figure 9). We thus conclude that unwanted metallic lead and associated phase impurities, which otherwise act as centres of degradation and carrier recombination, are stabilised and rendered electronically benign by conversion to wide bandgap PbO in the presence of oxygen and light.

The schematic in Figure 3k illustrates the proposed nanoscale mechanism taking place under illumination in oxygen. Molecular oxygen adsorbs to under-coordinated lead on the surface of the sample,[33] thus accumulating readily at highly defective regions. Photoexcited electrons reduce the adsorbed oxygen molecule yielding a superoxide ion which passivates surface trap states by covalently bonding at a halide vacancy site,[31] or reacting with an iodine interstitial to form iodate[34]. Surface-bound oxygen inhibits ionic rearrangement, which otherwise provides the halide 'fuel' for photo-chemical degradation, and thus inhibits iodine loss at grain surfaces. Further reduction of bound superoxide by interaction with photo-excited electrons creates peroxide;[35] PbO formation occurs via the reaction of this peroxide and the adjacent surface $Pb^{2+}$. PbO passivates the material surface by preventing accumulation of halide vacancies or metallic $Pb^0$, minimising parasitic carrier loss[31,36]. Occurring foremost at sites dense in under-coordinated Pb ions, the photo-chemical formation of PbO is specific to the defect-rich phase impurities and sites of initial degradation. Selective passivation at these sites results in more homogeneous optoelectronic properties after exposure to oxygen and light treatment.[37]

Our collective results show that in halide perovskite absorber layers the very sites of degradation are also sites of parasitic charge trapping, and are in nature localised phase impurities, including hexagonal polytypes and lead iodide. These residual phase impurities are not observable in macroscopic techniques in the highest performance films and are only revealed through local nanoscale measurements. Degradation seeds at phase impurities due to their high density of defects which act as both non-radiative recombination sites as well as fuel for redox photochemistry. Crucially, the relatively low defect populations in pristine sample regions inhibit the activity of otherwise sinister photochemical processes, though further work will need to elucidate whether any eventual degradation will be slow relative to the required operational lifetimes of optoelectronic devices such as solar cells and LEDs.[38] We conclude that urgent work must target the removal of even trace phase impurities from perovskite device layers, either at the film formation stage or with post-processing. The rewards will be twofold: reduced power losses through the removal of traps, and improved device longevity. Recent works have shown the addition of inorganic cations (Cs and/or Rb) to reduce the fraction of the polytype impurities,[39–41] though we find that even trace inclusions appearing sparsely in films will compromise stability such that this strategy is not sufficient. Additives themselves should be photo-stable and must not induce further charge trapping either initially or upon aging. $PbI_2$, which is often added to perovskite solar cells in excess to passivate grain boundaries and to obtain improved initial device performance,[42] does not satisfy this criteria; even if the $PbI_2$ appears in epitaxial form,[15] it still compromises long-term stability[29,43] and therefore must be eradicated from state-of-the-art device layers. Oxygen and light treatments provide a template for mitigating phase impurities by chemical conversion to benign species; although not a viable long-term solution due to the dynamic and reversible nature of the oxygen treatment, alternatives must be sought that mimic this process. In fact, oxygen here acts similarly to iodine blocking layers which have been introduced to stabilise perovskite solar cell modules.[44] Searching for other species capable of converting any unwanted products, such as phase impurities and metallic lead, to benign species should also be a pressing pursuit. The future of perovskite optoelectronics will depend on the absolute removal of phase impurities from cutting-edge devices, and thus the suppression of the redox photo-chemistry which will otherwise fatally limit operational lifetime.

## References

1. Stranks, S. D. & Snaith, H. J. Metal-halide perovskites for photovoltaic and light-emitting devices. *Nat. Nanotechnol.* **10**, 391–402 (2015).

2. NREL. Best Research-Cell Efficiency Chart. https://www.nrel.gov/pv/assets/pdfs/best-research-cell-efficiencies.20200406.pdf (2020).

3. Jošt, M., Kegelmann, L., Korte, L. & Albrecht, S. Monolithic Perovskite Tandem Solar Cells: A Review of the Present Status and Advanced Characterization Methods Toward 30% Efficiency. *Adv. Energy Mater.* **n/a**, 1904102.

4. Li, N., Niu, X., Chen, Q. & Zhou, H. Towards commercialization: the operational stability of perovskite solar cells. *Chem. Soc. Rev.* **49**, 8235–8286 (2020).

5. Doherty, T. A. S. *et al.* Performance-limiting nanoscale trap clusters at grain junctions in halide perovskites. *Nature* **580**, 360–366 (2020).

6. Miller, O. D., Yablonovitch, E. & Kurtz, S. R. Strong Internal and External Luminescence as Solar Cells Approach the Shockley–Queisser Limit. *IEEE J. Photovolt.* **2**, 303–311 (2012).

7. Tian, Y. *et al.* Mechanistic insights into perovskite photoluminescence enhancement: light curing with oxygen can boost yield thousandfold. *Phys. Chem. Chem. Phys.* **17**, 24978–24987 (2015).

8. Brenes, R. *et al.* Metal Halide Perovskite Polycrystalline Films Exhibiting Properties of Single Crystals. *Joule* **1**, 155–167 (2017).

9. Saliba, M. *et al.* Cesium-Containing Triple Cation Perovskite Solar Cells: Improved Stability, Reproducibility and High Efficiency. *Energy Env. Sci* **9**, 1989–1997 (2016).

10. Al-Ashouri, A. *et al.* Monolithic perovskite/silicon tandem solar cell with >29% efficiency by enhanced hole extraction. *Science* **370**, 1300–1309 (2020).

11. Andaji-Garmaroudi, Z., Anaya, M., Pearson, A. J. & Stranks, S. D. Photobrightening in Lead Halide Perovskites: Observations, Mechanisms, and Future Potential. *Adv. Energy Mater.* **n/a**, 1903109.

12. Tress, W. *et al.* Interpretation and evolution of open-circuit voltage, recombination, ideality factor and subgap defect states during reversible light-soaking and irreversible degradation of perovskite solar cells. *Energy Environ. Sci.* **11**, 151–165 (2018).

13. Lin, Y.-H. *et al.* A piperidinium salt stabilizes efficient metal-halide perovskite solar cells. *Science* **369**, 96–102 (2020).

14. Shao, S. & Loi, M. A. The Role of the Interfaces in Perovskite Solar Cells. *Adv. Mater. Interfaces* **7**, 1901469 (2020).

15. Rothmann, M. U. *et al.* Atomic-scale microstructure of metal halide perovskite. *Science* **370**, (2020).

16. Palosz, B. The structure of PbI$_2$ polytypes 2H and 4H: a study of the 2H-4H transition. *J. Phys. Condens. Matter* **2**, 5285–5295 (1990).

17. Meggiolaro, D. *et al.* Iodine chemistry determines the defect tolerance of lead-halide perovskites. *Energy Environ. Sci.* **11**, 702–713 (2018).

18. Motti, S. G. *et al.* Controlling competing photochemical reactions stabilizes perovskite solar cells. *Nat. Photonics* **13**, 532–539 (2019).

19. Samu, G. F. *et al.* Electrochemical Hole Injection Selectively Expels Iodide from Mixed Halide Perovskite Films. *J. Am. Chem. Soc.* **141**, 10812–10820 (2019).

20. Frolova, L. A. *et al.* Reversible Pb2+/Pb0 and I−/I3− Redox Chemistry Drives the Light-Induced Phase Segregation in All-Inorganic Mixed Halide Perovskites. *Adv. Energy Mater.* **11**, 2002934 (2021).

21. Azpiroz, J. M., Mosconi, E., Bisquert, J. & Angelis, F. D. Defect migration in methylammonium lead iodide and its role in perovskite solar cell operation. *Energy Environ. Sci.* **8**, 2118–2127 (2015).

22. deQuilettes, D. W. *et al.* Photo-induced halide redistribution in organic-inorganic perovskite films. *Nat Commun* **7**, 11683 (2016).

23. Wang, S., Jiang, Y., Juarez-Perez, E. J., Ono, L. K. & Qi, Y. Accelerated degradation of methylammonium lead iodide perovskites induced by exposure to iodine vapour. *Nat. Energy* **2**, 1–8 (2016).

24. Donakowski, A. *et al.* Improving Photostability of Cesium-Doped Formamidinium Lead Triiodide Perovskite. *ACS Energy Lett.* 574–580 (2021) doi:10.1021/acsenergylett.0c02339.

25. Baltog, I., Piticu, I., Constantinescu, M., Ghita, C. & Ghita, L. Optical investigations of PbI2 single crystals after thermal treatment. *Phys. Status Solidi A* **52**, 103–110 (1979).

26. Albrecht, M. G. & Green, M. The kinetics of the photolysis of thin films of lead iodide. *J. Phys. Chem. Solids* **38**, 297–306 (1977).

27. Juarez-Perez, E. J. *et al.* Photodecomposition and thermal decomposition in methylammonium halide lead perovskites and inferred design principles to increase photovoltaic device stability. *J. Mater. Chem. A* **6**, 9604–9612 (2018).

28. Cho, H. *et al.* Overcoming the electroluminescence efficiency limitations of perovskite light-emitting diodes. *Science* **350**, 1222–1225 (2015).

29. Tumen-Ulzii, G. *et al.* Detrimental Effect of Unreacted PbI2 on the Long-Term Stability of Perovskite Solar Cells. *Adv. Mater.* **32**, 1905035 (2020).

30. Feldmann, S. *et al.* Photodoping through local charge carrier accumulation in alloyed hybrid perovskites for highly efficient luminescence. *Nat. Photonics* 1–6 (2019) doi:10.1038/s41566-019-0546-8.

31. Godding, J. S. W. *et al.* Oxidative Passivation of Metal Halide Perovskites. *Joule* **3**, 2716–2731 (2019).

32. Szemjonov, A. *et al.* Impact of Oxygen on the Electronic Structure of Triple-Cation Halide Perovskites. *ACS Mater. Lett.* **1**, 506–510 (2019).

33. Huang, L., Ge, Z., Zhang, X. & Zhu, Y. Oxygen-induced defect-healing and photo-brightening of halide perovskite semiconductors: science and application. *J. Mater. Chem. A* **9**, 4379–4414 (2021).

34. Meggiolaro, D., Mosconi, E. & De Angelis, F. Mechanism of Reversible Trap Passivation by Molecular Oxygen in Lead-Halide Perovskites. *ACS Energy Lett.* **2**, 2794–2798 (2017).

35. Anaya, M., Galisteo-López, J. F., Calvo, M. E., Espinós, J. P. & Míguez, H. Origin of Light-Induced Photophysical Effects in Organic Metal Halide Perovskites in the Presence of Oxygen. *J. Phys. Chem. Lett.* **9**, 3891–3896 (2018).

36. Ouyang, Y. *et al.* Photo-oxidative degradation of methylammonium lead iodide perovskite: mechanism and protection. *J. Mater. Chem. A* **7**, 2275–2282 (2019).

37. Brenes, R., Eames, C., Bulović, V., Islam, M. S. & Stranks, S. D. The Impact of Atmosphere on the Local Luminescence Properties of Metal Halide Perovskite Grains. *Adv. Mater.* **30**, 1706208 (2018).

38. Meng, L., You, J. & Yang, Y. Addressing the stability issue of perovskite solar cells for commercial applications. *Nat. Commun.* **9**, 5265 (2018).

39. Gratia, P. *et al.* The Many Faces of Mixed Ion Perovskites: Unraveling and Understanding the Crystallization Process. *ACS Energy Lett.* **2**, 2686–2693 (2017).

40. Hu, Y. *et al.* Understanding the Role of Cesium and Rubidium Additives in Perovskite Solar Cells: Trap States, Charge Transport, and Recombination. *Adv. Energy Mater.* **8**, 1703057 (2018).

41. Dang, H. X. *et al.* Multi-cation Synergy Suppresses Phase Segregation in Mixed-Halide Perovskites. *Joule* **3**, 1746–1764 (2019).

42. Jacobsson, T. J. *et al.* Unreacted PbI$_2$ as a Double-Edged Sword for Enhancing the Performance of Perovskite Solar Cells. *J. Am. Chem. Soc.* **138**, 10331–10343 (2016).

43. Roose, B., Dey, K., Chiang, Y.-H., Friend, R. H. & Stranks, S. D. Critical Assessment of the Use of Excess Lead Iodide in Lead Halide Perovskite Solar Cells. *J. Phys. Chem. Lett.* **11**, 6505–6512 (2020).

44. Bi, E. *et al.* Efficient Perovskite Solar Cell Modules with High Stability Enabled by Iodide Diffusion Barriers. *Joule* **3**, 2748–2760 (2019).